# STUDY OF SAHARAN DUST INFLUENCE ON $PM_{10}$ MEASURES IN SICILY FROM 2013 TO 2015

A. Cuspilici, P. Monforte, M.A.Ragusa *

Department of Mathematics and Computer Science

Catania University

Corresponding author *maragusa@dmi.unict.it

*Abstract*

*Nowadays, particulate matter, especially that with small dimension as $PM_{10}$, $PM_{2.5}$ and $PM_1$, is the air quality indicator most commonly associated with a number of adverse health effects. In this paper it is analyzed the impact that a natural event, such as the transport of Saharan dust, can have on increasing the particulate matter concentration in Sicily. Consulting the data of daily $PM_{10}$ concentration, acquired by air quality monitoring network belonging to "Agenzia Regionale Protezione dell'Ambiente" (Environmental Protection Regional Agency), it was possible to analyze the trend from 2013 to 2015. The days, in which the limit value was exceeded, were subjected to combined analysis. It was based on three models: interpretations of the air masses back-trajectories, using the atmospheric model HYSPLIT (HYbrid Single-Particle Lagrangian Integrated trajectory); on the calculation of the concentration on the ground and at high altitude particulate applying DREAM model (Dust REgional atmospheric model) and on the calculation of the concentration of mineral aerosols according to the atmospheric optical thickness (AOT) applying NAAPS model (Navy Aerosol Analysis and Prediction System).The daily limit value exceedances were attributed to the transport of Saharan dust events exclusively when the three models were in agreement with each other. Identifying the natural events, it was possible to quantify the contribution of the Saharan dust and consequently the reduction of the exceedances number. To quantify the contribution of Saharan dust on daily $PM_{10}$ concentration, it was calculated the regional background in according to precautionary approach recommended by "Guidance on the quantification of the contribution of natural sources under the EU Air Quality Directive 2008/50/EC", when the application of the method cannot be validated with chemical analysis, as in this case. In this study is obtained, as the most important quantitative goal, the convergence of the three models to the same result. So, is evident that exceedances of the daily limit value that occurred from 2013 to 2015 in Sicily can be attributed, in most cases, to the Saharan dust intrusion.*

**1. Introduction**

The increasing modern agriculture and irrigation systems, which leads environmental issues as air pollutions, has been deeply studied on the last decade. We recall, for instance, the studies of Valipour, (2015a, 2015b, 2016a, 2016b, 2016c), Yannopoulos, et al. (2015).

The awareness of the public opinion and institutions to environmental issues in recent years encouraged the development of methodologies for the collection and analysis of air pollutants. Considering the amount of air daily inhaled by an individual (from 6 to 9 l/min at rest condition, 60 l/min during moderate exercise), it is possible to better realize the importance for the health and the risks associated with polluted air respiration.

Nowadays, in the major Italian and European cities, the most evenly dispersed pollutant is particulate matter Künzli et al.,(1999), EEA, (2015), Capizzi et al. (2015). Suspended particulates, especially those ones with small aerodynamic diameter such as $PM_{10}$, $PM_{2.5}$, $PM_1$, are the most commonly air quality indicators associated with several adverse health effects, Caramagna et al., (2015), Strano et al., (2015). Due to their intrinsic characteristics, they are able to remain in atmosphere for a longer time, representing so a danger to the health of humans, animals and plants. Epidemiological studies done by Research on Cancer showed that the particulate increase facilitates the appearance of chronic lung disease; the formation of various malignancies and rises the mortality from cardiovascular and respiratory diseases. Studies done during the last decade showed an association between levels of air pollutants and the daily number of deaths or hospital admissions for respiratory and cardiovascular causes, Glinianaia, et al., (2004), Maisonet, et al., (2004). The possible damage for the human body may derive from the type of particle which could be toxic by itself, or more frequently, due to the substances deposited on it. Particulate matter can be considered the means which allows the penetration into the respiratory system of human body, of potentially harmful substances, Schwartz, et al., (1996), Norris et al., (1999), Pope, (2000a, 2000b), Ostro, et al., (2006), Dockery and Stone , (2007), Kampa and Castanas, (2008). In addition to the damage on health, particulate matter has a very important role on global climate change, Charlson, (1992), Duro, (2014, 2015). It has effects in the propagation and absorption of solar radiation, on the atmospheric visibility and in the processes of vapor condensation, favoring the formation of smog and fog, Wexler and Ge Z., (1998), Kaufman, (2006). Although the pollution originated by human activities is the most accused in the deterioration of air quality, pollution of natural origin such as Saharan dust intrusion must not be underestimated. Concerning the specific health effects connected to Saharan dust intrusion, there are not firm and adequately shared evidences yet, further investigations aimed to understand the role of natural events in air pollution and the consequent effects on human health are needed, Linares et al., (2010).

Generally, these intrusions can cause abnormal rise in $PM_{10}$ concentration values, and, in some cases, contribute to excess the limit values fixed by European Directive 2008/50/EC. European Directive requires that, in all Member States the average daily concentration of $PM_{10}$ should not exceed, more than 35 times during a calendar year, the limit value of 50 µg/m$^3$. However, it allows to separate exceedances due to natural contributions, if this natural occurrence is demonstrated with sufficient certainty. The guidelines issued by the European Commission in 2011 established the basic elements for the identification and evaluation of the desert sands impact on $PM_{10}$ levels measured in air quality monitoring stations, European Commission (2011), Marelli, (2007). Consulting the data of daily $PM_{10}$ concentrations, acquired by three measuring stations belonging to the air quality monitoring network of the *Agenzia Regionale Protezione dell'Ambiente* (Environmental Protection Regional Agency), it was possible to analyze the trend from 2013 to 2015. The days in which limit value exceedances were recorded, they were submitted to combined analysis. This analysis is based on interpretations of the back-trajectories of air masses, which was obtained by HYSPLIT model (Hybrid Single Particle Lagrangian Integrated Trajectory), Draxler and Hess (1997), Draxler et al., (2010); the calculation of the concentration of particulate matter in soil and altitude was obtained applying DREAM model (Dust Regional Atmospheric Model), Basart et al., (2012); while the calculation of the concentration of mineral aerosols as a function of the atmospheric optical thickness (AOT) was obtained using NAAPS model (Navy Aerosol Analysis and Prediction System), Jaffe et al., (2004). Exceedances of the daily limit value (DLV) were attributed to the transport of Saharan dust events when the three models (simultaneously verified) applied during the days gave the same results. Identifying the natural events, it was possible to quantify the contribution of the Saharan dust and consequently to reduce, where possible, the number of exceedances recorded from 2013 to 2015 in the three monitoring stations considered.In this work is analyzed the impact of a natural event, such as the transport of Saharan dust, which may have on the increase in particulate matter concentration in Sicily. The novelty of this investigation is the approach followed by guidelines, which provides for the combined analysis of the weather conditions, the model forecasts, satellite observations, measurements of $PM_{10}$ and the suggestion of a working methodology for the assessment of transport events sand applicable in each zone that is object of interest, see e.g. Pey et al., (2013), Querol et al., (2009). So, it is possible to extend the results to other regions with similar conditions. In this study the authors did not measure $PM_{2.5}$ because the analyzer present in the monitoring stations does not allow to acquire these concentrations. The authors intend to improve the research in this direction.

## 2. Measuring of daily $PM_{10}$ concentration

The daily data of $PM_{10}$ concentrations, originating from the three stations which belong to the air quality monitoring regional network were analyzed. The monitoring stations were selected in purpose of the climatic variability that affects the region. In detail, the monitoring station identified as Partinico (PTN) represents Western Sicily, the monitoring station identified as Enna (ENN) represents Central Sicily and the monitoring station identified as Misterbianco (MRB) represents Eastern Sicily (figure 1 and table 1). Such monitoring stations adopt the β-ray attenuation method, for the measurements of the mass concentrations of $PM_{10}$. According to the Directive 2008/50/EC the daily concentration of $PM_{10}$ must not exceed, more than 35 times during a calendar year, the limit value of 50 µg/m$^3$, while the annual average should not exceed 40 µg/m$^3$. In order to verify compliance with the limit values, it was analyzed the trend of the daily $PM_{10}$ concentrations detected in the three monitoring stations from 2013 to 2015 (figures 2, 3, 4). The trends have shown compliance with the limit values prescribed by the regulations for the three years analyzed. Moreover it was possible to identify days with anomalous $PM_{10}$ concentrations. In the majority of cases the abnormal concentrations have affected the three measurement sites simultaneously. This has helped to narrow the hypotheses about possible causes excluding, for example, the urban traffic and assuming instead a Saharan dust event.

## 3. Identification of Saharan dust intrusion

The guidelines issued by European Commission set out the basic elements for identification and evaluation of impact of desert dust on $PM_{10}$ levels measured in air quality monitoring stations, without requiring knowledge about character particulate. In order to identify the episodes of Saharan dust transport in Sicily, a combined analysis based on weather conditions, on model predictions and action on the ground was conducted. The methodology applied in this work takes account of the following:

- Interpretation of the daily weather conditions by analyzing the back-trajectories of air masses, with HYSPLIT model (HYbrid Single-Particle Lagrangian Integrated Trajectory) developed by National Oceanic and Atmospheric Administration, USA (NOAA), Escudero et al., (2006), Stein et al., (2015). HYSPLIT accomplishes a wide range of simulations related to atmospheric transport and dispersion of pollutants using meteorological archived data and forecasts, provides information on the trajectories of air masses that can carry particulate. As specified in the guidelines [16] the back-trajectories were calculated starting from 12:00 a.m. of the day when there was an overrun and working backward for five days, to heights of 750, 1500, 2500 meters above sea level. About the receptors, the geographical coordinates of the three monitoring stations have been used.

- Application of DREAM model (Dust REgional Atmospheric Model) developed by Barcelona Supercomputing Center, Pérez et al., (2006a). The model predicts, each six hours, the life cycle in the atmosphere of the particulate eroded from desert areas, providing information on the concentration to the ground and at high altitude.

- Application of NAAPS model (Navy Aerosol Analysis and Prediction System), developed by Naval Research Laboratory (http://www.nrlmry.navy.mil/aerosol/). The model is capable of estimating concentrations of mineral aerosols, defining the origin as a function of the atmospheric optical thickness (AOT), NAVY, US., (2003), Pérez et al., (2006b). Greater is optical thickness for a wavelength; shorter is the amount of light of that wavelength that reaches the earth's surface. This model provides, for each day analyzed with an hourly resolution of six hours, the evolution of any clouds of mineral aerosols.

The days when the concentration values measured in the three monitoring stations have exceeded the DVL of 50 $\mu g/m^3$ were subjected to this procedure. The exceedance of DVL has been attributed to the Saharan dust transport exclusively in the case in which the three models were in agreement with each other, for the day in question. In figures 5,6,7 it is shown the agreement between the three models during a day in which DVL exceed; this overrun was later attributed to the intrusion of Saharan dust. Figure 5 shows the back-trajectories calculated by HYSPLIT model in three receptor sites. It is evident that during the five days preceding the event, air masses tend to move from North Africa to Sicily.
Figure 6 shows the dust concentrations at the ground obtained by DREAM model. It provides a concentration range between 80-160 $\mu g/m^3$. The concentration values measured by MRB and PTN located at opposite sides of the island, were respectively 137.3 $\mu g/m^3$ and 76.2 $\mu g/m^3$. The concentration measured in MRB is within the range predicted by model. The concentration measured in PTN is instead lower by 5% compared to the lower limit of the range. The interval calculated by model represents satisfactorily the real conditions of the soil concentrations.
Figure 7 shows the extent to which the aerosols affect the passage of sunlight through the atmosphere. Measurements of aerosols optical thickness (AOT) in more than one wavelength are able to provide important information on the concentration, type and variability of aerosols in the atmosphere. For the day analyzed the NAAPS model identifies a cloud of mineral aerosols (band visible from yellow to green) that covering the lower part of the Mediterranean basin, affecting in particular North Africa and Sicily. In addition it can be seen as in Eastern Sicily the optical thickness is greater than the one found in the Western area, according to $PM_{10}$ concentration measured in the respective monitoring stations (MRB and PTN).
According to the results obtained from the models, it was possible to identify events of Saharan dust and consequently reasons DLV exceeding measured at the monitoring stations. This methodology allowed to recognize the days when the DVL overruns are attributable to anthropogenic causes, from those attributable to the Saharan dust events that affected Sicily from 2013 to 2015.

## 4. Contribution of Saharan dust

The exceedance days which can be attributable to Saharan dust events and measured in the reference stations, were identified. To quantify the contribution of Saharan dust on $PM_{10}$ daily concentration, regional background was calculated, Escudero ed al., (2007). It is represented by the 50[th] percentile of the $PM_{10}$ series values measured 15 days before and 15 days after the Saharan dust event. It was used 50[th] percentile to estimate the regional background according to the precautionary approach recommended by *Guidance on the quantification of the contribution of natural sources under the EU Air QualityDirective 2008/50/EC* when the application methodology cannot be validated with chemical analysis, as in this case. Therefore, the contribution of Saharan dust was calculated by (i):

$$C_{ij\ DUST} = Dc_{ij} - RB_{ij} \quad (i)$$

where:

$C_{ij\ DUST}$ = Saharan dust contribution measured in an $i^{th}$ site during a $j^{th}$ event day [µg/m$^3$]
$Dc_{ij}$ = PM$_{10}$ daily concentration measured in an $i^{th}$ site during a $j^{th}$ event day [µg/m$^3$]
$RB_{ij}$ = regional background for $i^{th}$ site calculated for a $j^{th}$ event day [µg/m$^3$]

DLV exceedance is canceled if $C_{ij\ DUST}$ is less than 50 µg/m$^3$. The subtraction of the contribution from the Saharan dust to PM$_{10}$ measures has reduced the number of exceedances from 2013 to 2015 in Sicily. The percentage of reduction (R) about number of exceedances of the DLV was calculated by (ii):

$$R = \frac{A_E - E_{AS}}{A_E} \cdot 100 \quad (ii)$$

where:

$A_E$ = total number of exceedances measured during the year;
$E_{AS}$ = number of exceedances after subtracting the contribution of Saharan dust from the measuring.

Table 2 summarizes the number of annual exceedances observed in the three monitoring stations, the number of exceedances after subtraction Saharan dust contribution and the percentage of DLV overruns reduction. Figure 8 shows the graphical analysis of the percentage reduction about number of exceedances from 2013 to 2015 for the three monitoring stations.

## 5. Conclusions

Transport of Saharan dust is a natural event extensively studied from different perspective, recognizing it a significant role on the different effects it produces on the climate and the environment. The purpose of this work was to evaluate the impact of Saharan dust in Sicily on PM$_{10}$ concentrations. From PM$_{10}$ daily concentration, collected from three monitoring stations belonging to "*Agenzia Regionale per la Protezione dell'Ambiente*" (Enviromental Protection Regional Agency) it was possible to analyse the daily trend from 2013 to 2015. The days when concentrations exceeded the DLV of 50 μg/m$^3$, were subjected to a combined analysis based on concentration values at the ground, on the atmospheric optical thickness present in the Mediterranean basin during the days in question and on the reconstruction of the back-trajectories of air masses in order to track origin direction. The reconstruction of back-trajectories that passed through Sicily during the days of Saharan events were obtained by applying HYSPLIT model (HYbrid Single Particle Lagrangian Integrated Trajectory) developed by National Oceanic and Atmospheric Administration, USA (NOAA). The evaluation of mineral aerosol concentrations in the atmospheric optical thickness was obtained applying NAAPS model (Navy Aerosol Analysis and Prediction System), developed by Naval Research Laboratory. For the concentrations on the ground it was taken into account the forecast model DREAM (Dust REgional Atmospheric Model) developed by Barcelona Supercomputing Centre. When the three models agreed (during the days in which this procedure was applied), it was possible to identify the transport events of Saharan dust and therefore to proceed to the quantification of their contribution in PM$_{10}$ concentrations, as provided by *Guidance on the quantification of the contribution of natural sources under the EU, Air Quality Directive 2008/50/EC*. From the results it was evident that the DLV exceedances occurred from 2013 to 2015 in Sicily could be attributed, in the majority cases, to the Saharan dust intrusion. From the comparison of annual trends of PM$_{10}$ concentrations was possible to highlight the presence of outliers simultaneously in Western Sicily, Central and Eastern. The contemporaneity of the abnormal values of PM$_{10}$ concentrations is a further demonstration of how Sicily is strongly influenced by the intrusion of Saharan dust. The percentage of reduction of the days exceeding the DLV obtained for the three years analysed it has provided a measure of the phenomenon.
Future developments will focus on the chemical speciation of PM$_{10}$ in particular on concentrations of crustal components (Ca, Mg, Fe, Si and K), in correspondence of transport events of Saharan dust, in order to more accurately validate the identification and their quantification.

*Table 1. Identification of monitoring stations*

| Station | ID | Location | Altitude (m.a.s.l.) | $PM_{10}$ monitoring method |
|---|---|---|---|---|
| PA-Partinico | PTN | 38°02'58.9'' N 13°07'07.2'' E | 182 | β-Attenuation |
| EN-Enna | ENN | 37°33'51.8'' N 14°16'54.5'' E | 920 | β-Attenuation |
| CT-Misterbianco | MRB | 37°30'56.4'' N 15°00'38.5'' E | 177 | β-Attenuation |

*Table 2. Number of exceedances and percentage reduction for years and monitoring stations.*

| Monitoring station | Year | Annual Exceedances ($A_E$) | Exceedances After Subtraction ($E_{AS}$) | Reduction (%) |
|---|---|---|---|---|
| PTN | 2013 | 4 | 1 | 75 |
|  | 2014 | 20 | 8 | 60 |
|  | 2015 | 9 | 1 | 88.9 |
| ENN | 2013 | 4 | 3 | 25 |
|  | 2014 | 9 | 8 | 11.1 |
|  | 2015 | 6 | 1 | 83.3 |
| MRB | 2013 | 4 | 1 | 75 |
|  | 2014 | 16 | 7 | 56.2 |
|  | 2015 | 5 | 3 | 40 |

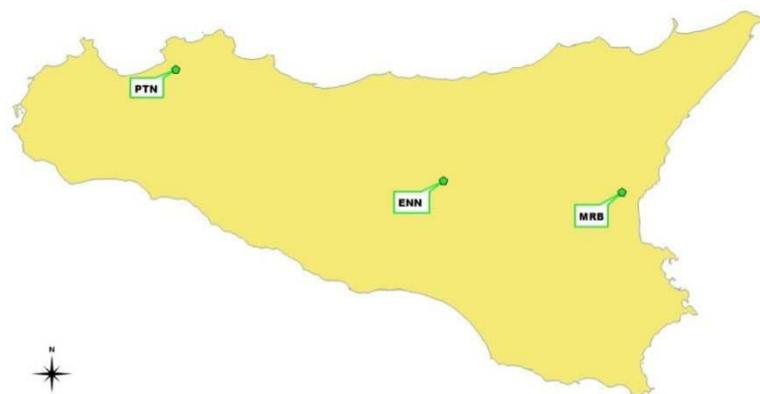

*Figure 1. Location of monitoring stations*

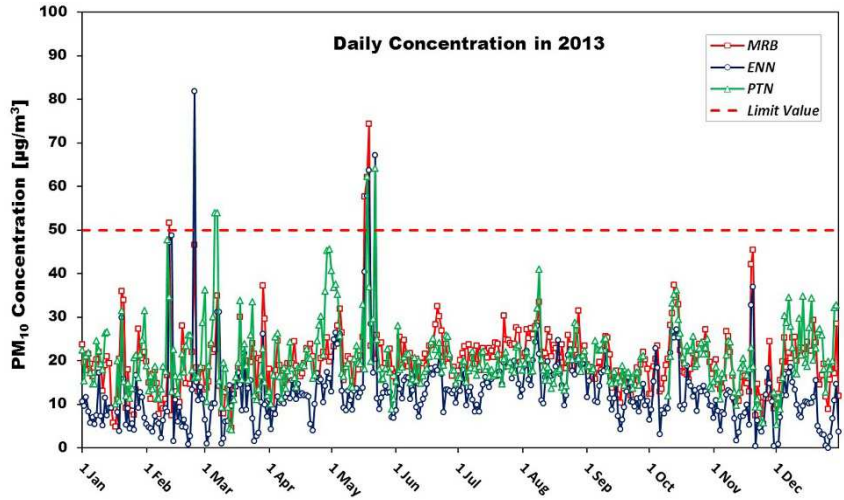

*Figure 2. Trend of daily concentration of PM$_{10}$ during 2013*

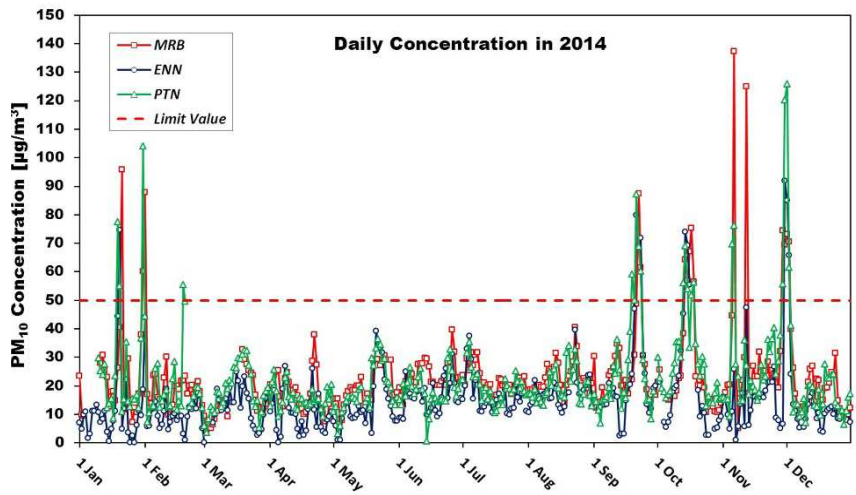

*Figure 3. Trend of daily concentration of PM$_{10}$ during 2014*

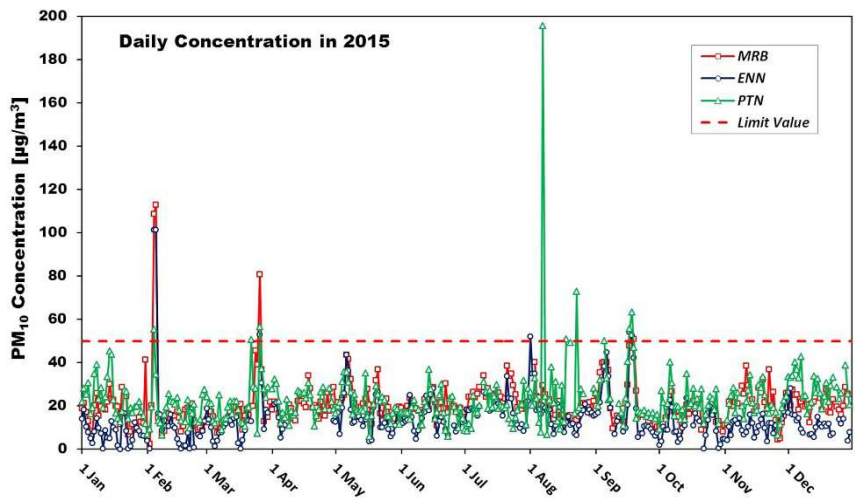

*Figure 4. Trend of daily concentration of PM$_{10}$ during 201*

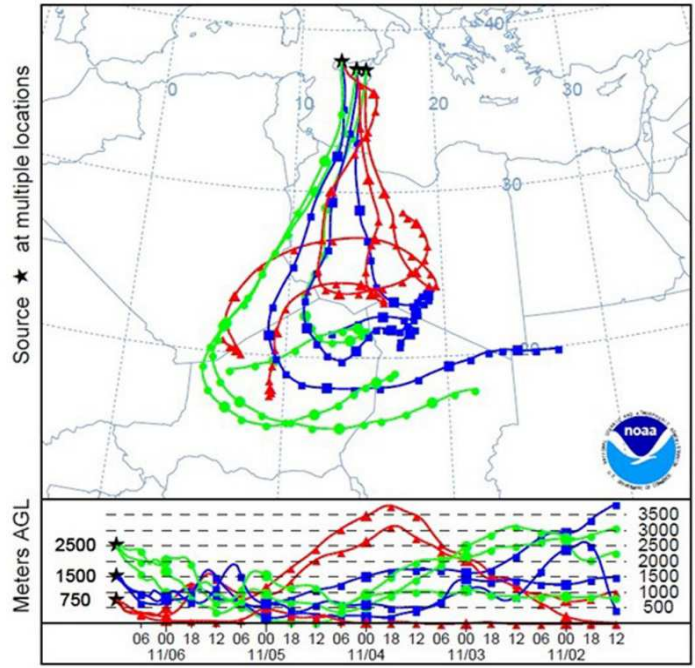

*Figure 5. Back trajectories by HYSPLIT model, five days before the event on 6$^{th}$ November 2014. The receptor positions represent the three monitoring stations. Air masses move from North Africa to Sicily.*

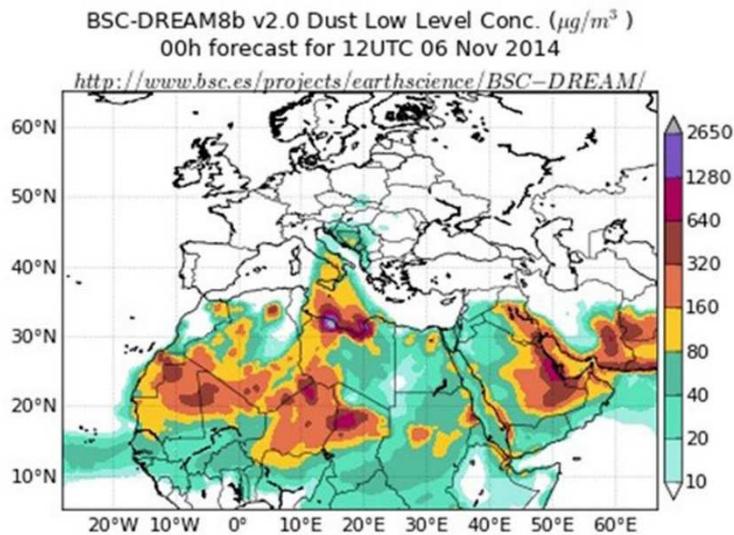

*Figure 6. Map of ground concentrations by DREAM model for the Mediterranean at 12:00 a.m. on 6$^{th}$ November 2014. In Sicily, the concentration range was between 80-160 µg/m$^3$*

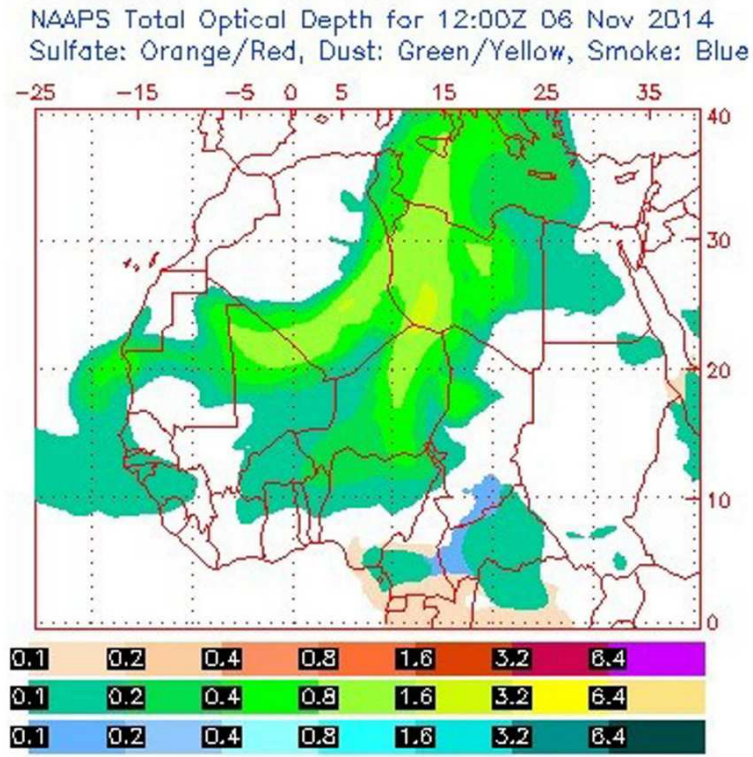

*Figure 7. Map by NAAPS model about extent to which aerosols affect the passage of sunlight through the atmosphere at 12:00a.m. on 6$^{th}$ November 2014. It is clearly visible the "cloud" of mineral aerosols (yellow-green) that it is stretching from North Africa to southern Europe affecting especially Sicily.*

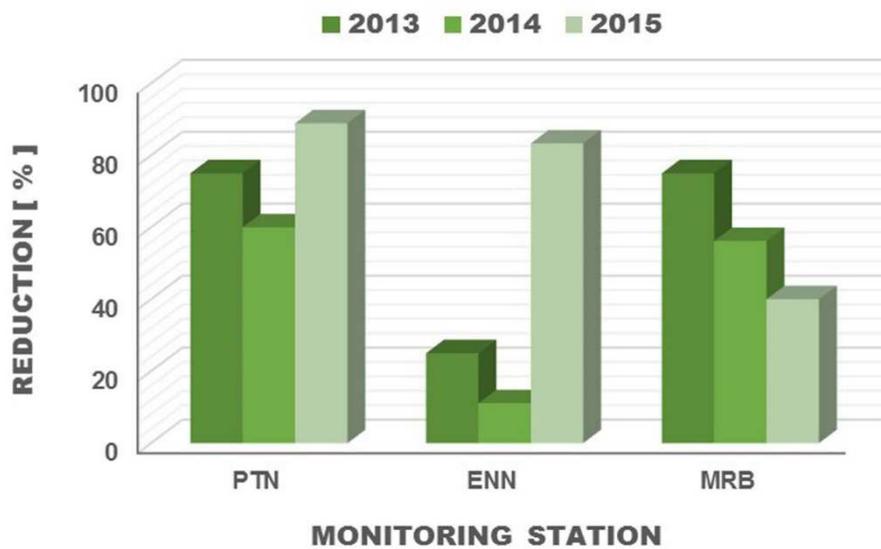

Figure 8. *Percentage reduction of exceedances from 2013 to 2015*